\documentclass[conference]{IEEEtran} 
\IEEEoverridecommandlockouts
\usepackage{cite}
\usepackage{amsmath,amssymb,amsfonts}
\usepackage{algorithmic}
\usepackage{graphicx}
\usepackage{textcomp}
\usepackage{xcolor}
\usepackage{url}
\def\BibTeX{{\rm B\kern-.05em{\sc i\kern-.025em b}\kern-.08em
    T\kern-.1667em\lower.7ex\hbox{E}\kern-.125emX}}

\begin{document}

\title{What is Normal? A Big Data Observational Science Model of Anonymized Internet Traffic
\thanks{
Research was sponsored by the Department of the Air Force Artificial Intelligence Accelerator and was accomplished under Cooperative Agreement Number FA8750-19-2-1000. The views and conclusions contained in this document are those of the authors and should not be interpreted as representing the official policies, either expressed or implied, of the Department of the Air Force or the U.S. Government. The U.S. Government is authorized to reproduce and distribute reprints for Government purposes notwithstanding any copyright notation herein. Use of this work is controlled by the human-to-human license listed in Exhibit 3 of https://doi.org/10.48550/arXiv.2306.09267
}
}

\author{\IEEEauthorblockN{
Jeremy Kepner$^1$, Hayden Jananthan$^1$, Michael Jones$^1$,  William Arcand$^1$, David Bestor$^1$, William Bergeron$^1$, \\ Daniel Burrill$^1$,  Aydin Buluc$^2$, Chansup Byun$^1$, Timothy Davis$^3$, Vijay Gadepally$^1$, Daniel Grant$^4$, Michael Houle$^1$, \\ Matthew Hubbell$^1$,  Piotr Luszczek$^{1,5}$, Lauren Milechin$^1$, Chasen Milner$^1$, Guillermo Morales$^1$, Andrew Morris$^4$, \\ Julie Mullen$^1$, Ritesh Patel$^1$, Alex Pentland$^1$, Sandeep Pisharody$^1$, Andrew Prout$^1$,  Albert Reuther$^1$, Antonio Rosa$^1$, \\ Gabriel Wachman$^1$, Charles Yee$^1$, Peter Michaleas$^1$
\\
\IEEEauthorblockA{
$^1$MIT,  $^2$LBNL, $^3$Texas A\&M, $^4$GreyNoise, $^5$University of Tennessee
}}}
\maketitle

\begin{abstract}
Understanding what is normal is a key aspect of protecting a domain.  Other domains invest heavily in observational science to develop models of normal behavior to better detect anomalies.
Recent advances in high performance graph libraries, such as the GraphBLAS, coupled with supercomputers enables processing of the trillions of observations required.
We leverage this approach to synthesize low-parameter observational models of anonymized Internet traffic with a high regard for privacy.
\end{abstract}

\begin{IEEEkeywords}
Internet traffic, anonymized analysis, streaming graphs, traffic matrices, network models
\end{IEEEkeywords}

\section{Introduction}

Anomaly detection and signature detection both play important roles in detecting adversarial activities on the Internet and both approaches are increasingly being enabled by big data and machine learning techniques \cite{ahmed2016survey, fernandes2019comprehensive, kwon2019survey, singla2019deep, kaur2020hybrid, wang2021machine}.
A core challenge to creating effective anomaly detection systems is the development of adequate models
\cite{chandola2009anomaly} 
of typical activity
\begin{quote}
{\small
\emph{The concept of normality} It is one of the main steps to build a solution to detect network anomalies. The question ``how to create a precise idea of normality?'' is what has driven most researchers into creating different solutions through the years. This can be considered as the main challenge related to anomaly detection and has not been entirely solved yet.
\cite{fernandes2019comprehensive}}
\end{quote}
Other domains (land, sea, undersea, air, and space) rely on detailed observational science models of their environment to understand what is normal
\cite{watson1957three, delaney1990air, delaney2015perspectives, geul2017modelling, o2018radar, pisharody2021realizing}.
 Accordingly, reproducible observations of cyberspace
\cite{jason2010science, carroll2012realizing, thuraisingham2016data, spring2017practicing}
have been recommended as a core foundation  for the science of cyber-security
\begin{quote}
{\small
The highest priority should be assigned to establishing research protocols to enable reproducible [observations].
\cite{jason2010science}}
\end{quote}
Significant early results from analyzing the Internet helped establish the emerging field of Network Science  
\cite{leland1994self, faloutsos1999power, barabasi1999emergence, albert1999internet, clauset2009power, mahanti2013tale, barabasi2016network, cao18impact}.
Improving these results requires ever larger data sets.  A priority for expanded observation of the Internet is the need to maintain a high regard for privacy.

\begin{figure}
\center{\includegraphics[width=1.0\columnwidth]{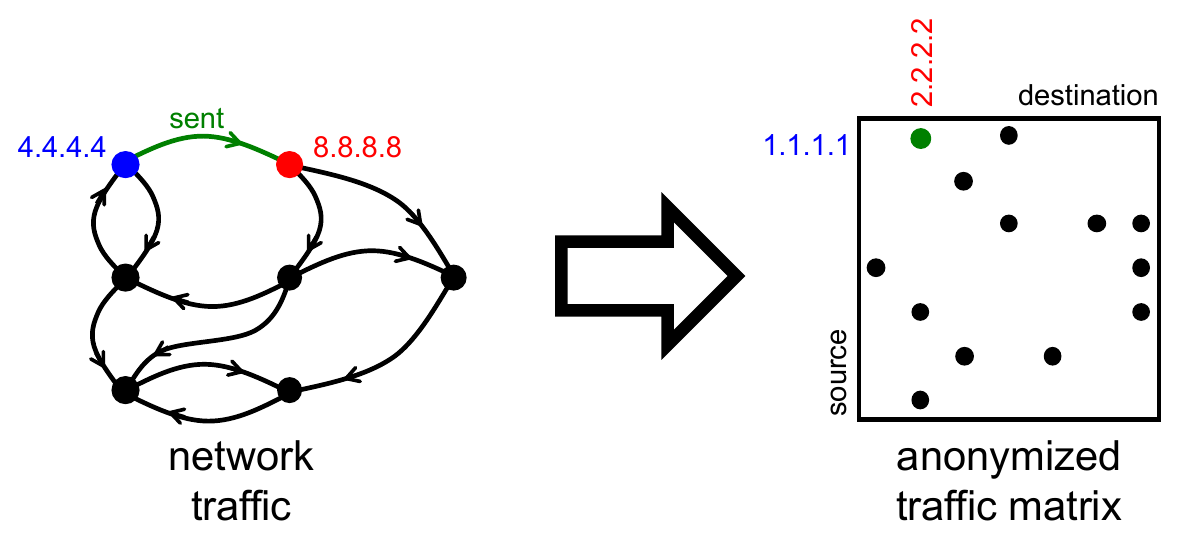}}
      	\caption{{\bf Network Traffic Messages to Anonymized Traffic Matrix.}   Network traffic uses numbers to the denote source  and destination addresses of messages.   Network traffic messages can be aggregated and summarized into traffic matrices for analysis.  These traffic matrices, when coupled with data sharing agreements, can be anonymized by relabeling source addresses (e.g., {\sf\textcolor{blue}{4.4.4.4}} $\rightarrow$ {\sf\textcolor{blue}{1.1.1.1}}) and destination addresses (e.g., {\sf\textcolor{red}{8.8.8.8}} $\rightarrow$ {\sf\textcolor{red}{2.2.2.2}})  using various anonymization schemes.}
      	\label{fig:NetworkToMatrix}
\end{figure}

The Center for Applied Internet Data Analysis (CAIDA)  based at the UC San Diego Supercomputer Center operates the largest Internet telescope in the world and has pioneered trusted data sharing best practices that combine anonymization \cite{fan2004prefix}
with data sharing agreements. These data sharing best practices include the following principles \cite{kepner2021zero}
\begin{itemize}
\item Data is made available in curated repositories
\item Standard anonymization methods are used where needed
\item Recipients register with the repository and demonstrate a legitimate research need
\item Recipients legally agree to neither repost a corpus nor deanonymize data
\item Recipients can publish analysis and data examples necessary to review research
\item Recipients agree to cite the repository and provide publications back to the repository
\item Repositories can curate enriched products developed by researchers
\end{itemize}

In the broader networking community (commercial, federal, and academia) anonymized source-to-destination traffic matrices with standard data sharing agreements have emerged as a data product that can meet many of these requirements (see Figure~\ref{fig:NetworkToMatrix})
\cite{uhlig2006providing, tune2013internet, vinayakumar2017applying, kepner2021zero}.
Focusing on anonymized source and destination addresses has helped alleviate privacy concerns because the non-anonymized addresses of Internet packets are already handled by many entities as part of the normal functioning of the Internet.

While an anonymized traffic matrix provides very little information about individual communications on a network, the ability collect trillions of observations over years across the Internet provides a unique opportunity for developing high-precision, low-parameter observational models of anonymized Internet traffic while maintaining a high regard for privacy. These data are particularly useful for addressing the fundamental observational science question for determining what is normal in a domain: \emph{Given two observers at different locations and/or times what can they both expect to see?}

The organization of the rest this paper is as follows.  First, some fundamental network quantities are presented along with how these quantities can be readily computed from anonymized traffic matrices.  Second, the statistical properties of these network quantities are examined from of largest available network data sets in the world (CAIDA, MAWI, GreyNoise, ...).  Next, these statistical results are synthesized into a low-parameter model of anonymized Internet traffic.  Finally, the papers concludes with a summary and a discussion of potential future directions.

\section{Traffic Matrices and Network Quantities}

\begin{figure}
\center{\includegraphics[width=1.0\columnwidth]{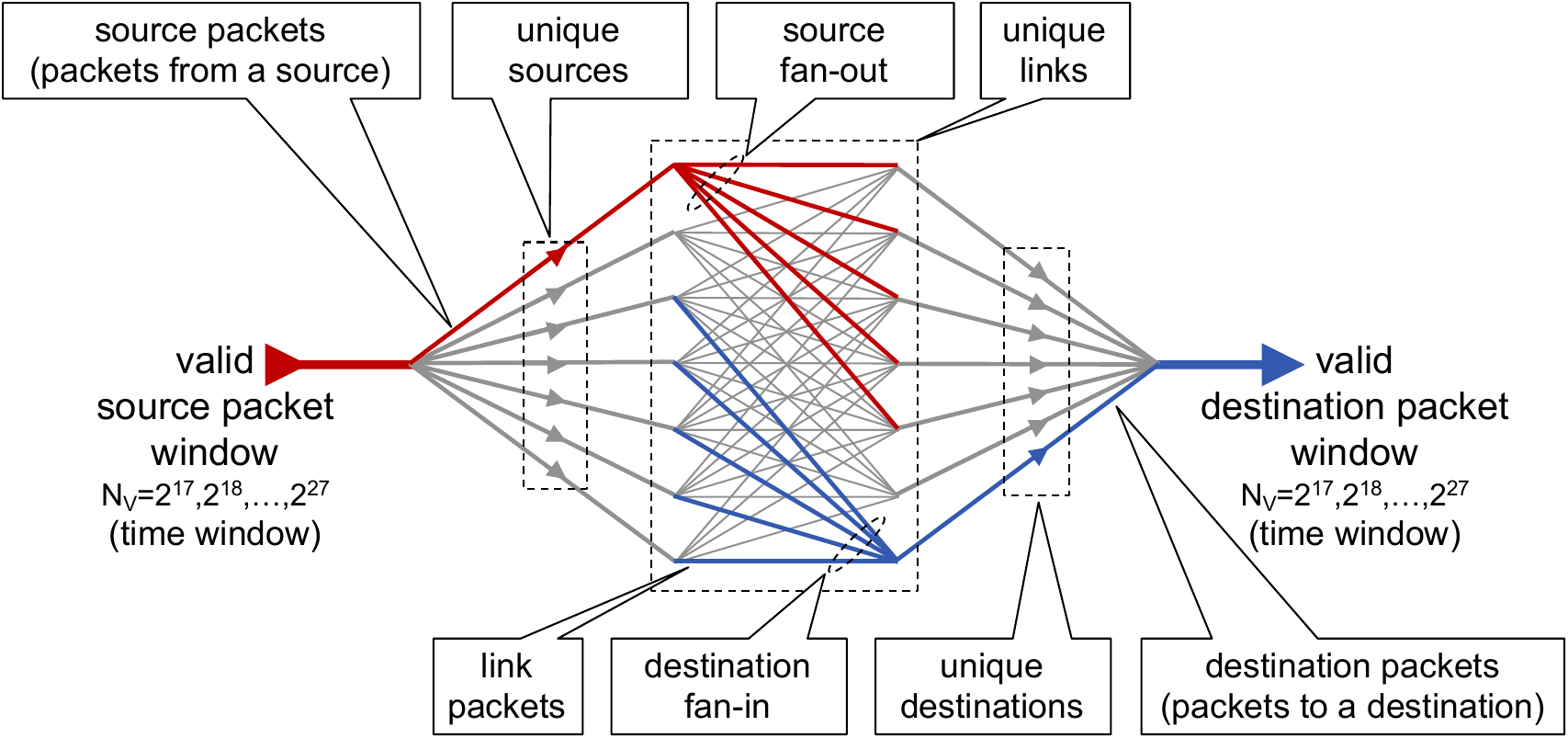}}
      	\caption{{\bf Streaming Network Traffic Quantities.} Internet traffic streams of $N_V$ valid packets are divided into a variety of quantities for analysis: source packets, source fan-out, unique source-destination pair packets (or links), destination fan-in, and destination packets.  Figure adapted from \cite{kepner19hypersparse}.}
      	\label{fig:NetworkDistribution}
\end{figure}

Network traffic data can be viewed as a traffic matrix where each row is a source and each column is a destination (see Figure~\ref{fig:NetworkToMatrix}). A primary benefit of constructing anonymized  traffic matrices with high performance math libraries, such as, the GraphBLAS\cite{davis2019algorithm}, is the efficient computation of a wide range of network quantities via matrix mathematics that enable trillions of events to be readily processed with supercomputers
\cite{gadepally2018hyperscaling}.
  Figure~\ref{fig:NetworkDistribution} illustrates essential quantities found in all streaming dynamic networks. These quantities are all computable from anonymized traffic matrices created from the source and destinations found in Internet packet headers.

\begin{table}
\caption{Network Quantities from Traffic Matrices}
\vspace{-0.25cm}
Formulas for computing network quantities from  traffic matrix ${\bf A}_t$ at time $t$ in both summation and matrix notation. ${\bf 1}$ is a column vector of all 1's, $^{\sf T}$  is the transpose operation, and $|~|_0$ is the zero-norm that sets each nonzero value of its argument to 1.  These formulas are unaffected by matrix permutations and will work on anonymized data.  Table adapted from \cite{kepner2020multi}.
\begin{center}
\begin{tabular}{p{1.45in}p{0.9in}p{0.6in}}
\hline
{\bf Aggregate} & {\bf ~~~~Summation} & {\bf ~Matrix} \\
{\bf Property} & {\bf ~~~~~~Notation} & {\bf Notation} \\
\hline
Valid packets $N_V$ & $~~\sum_i ~ \sum_j ~ {\bf A}_t(i,j)$ & $~{\bf 1}^{\sf T} {\bf A}_t {\bf 1}$ \\
Unique links & $~~\sum_i ~ \sum_j |{\bf A}_t(i,j)|_0$  & ${\bf 1}^{\sf T}|{\bf A}_t|_0 {\bf 1}$ \\
Link packets from $i$ to $j$ & $~~~~~~~~~~~~~~{\bf A}_t(i,j)$ & ~~~$~{\bf A}_t$ \\
Max link packets ($d_{\rm max}$) & $~~~~~\max_{ij}{\bf A}_t(i,j)$ & $\max({\bf A}_t)$ \\
\hline
Unique sources & $~\sum_i |\sum_j ~ {\bf A}_t(i,j)|_0$  & ${\bf 1}^{\sf T}|{\bf A}_t {\bf 1}|_0$ \\
Packets from source $i$ & $~~~~~~~\sum_j ~ {\bf A}_t(i,j)$ & ~~$~~{\bf A}_t  {\bf 1}$ \\
Max source packets ($d_{\rm max}$)  & $ \max_i \sum_j ~ {\bf A}_t(i,j)$ & $\max({\bf A}_t {\bf 1})$ \\
Source fan-out from $i$ & $~~~~~~~~~~\sum_j |{\bf A}_t(i,j)|_0$  & ~~~$|{\bf A}_t|_0 {\bf 1}$ \\
Max source fan-out ($d_{\rm max}$) & $ \max_i \sum_j |{\bf A}_t(i,j)|_0$  & $\max(|{\bf A}_t|_0 {\bf 1})$ \\
\hline
Unique destinations & $~\sum_j |\sum_i ~ {\bf A}_t(i,j)|_0$ & $|{\bf 1}^{\sf T} {\bf A}_t|_0 {\bf 1}$ \\
Destination packets to $j$ & $~~~~~~~\sum_i ~ {\bf A}_t(i,j)$ & ${\bf 1}^{\sf T}|{\bf A}_t|_0$ \\
Max destination packets ($d_{\rm max}$) & $ \max_j \sum_i ~ {\bf A}_t(i,j)$ & $\max({\bf 1}^{\sf T}|{\bf A}_t|_0)$ \\
Destination fan-in to $j$ & $~~~~~~~~~~\sum_i |{\bf A}_t(i,j)|_0$ & ${\bf 1}^{\sf T}~{\bf A}_t$ \\
Max destination fan-in ($d_{\rm max}$) & $ \max_j \sum_i |{\bf A}_t(i,j)|_0$ & $\max({\bf 1}^{\sf T}~{\bf A}_t)$ \\
\hline
\end{tabular}
\end{center}
\label{tab:Aggregates}
\end{table}%

\begin{figure*}
\center{\includegraphics[width=1.8\columnwidth]{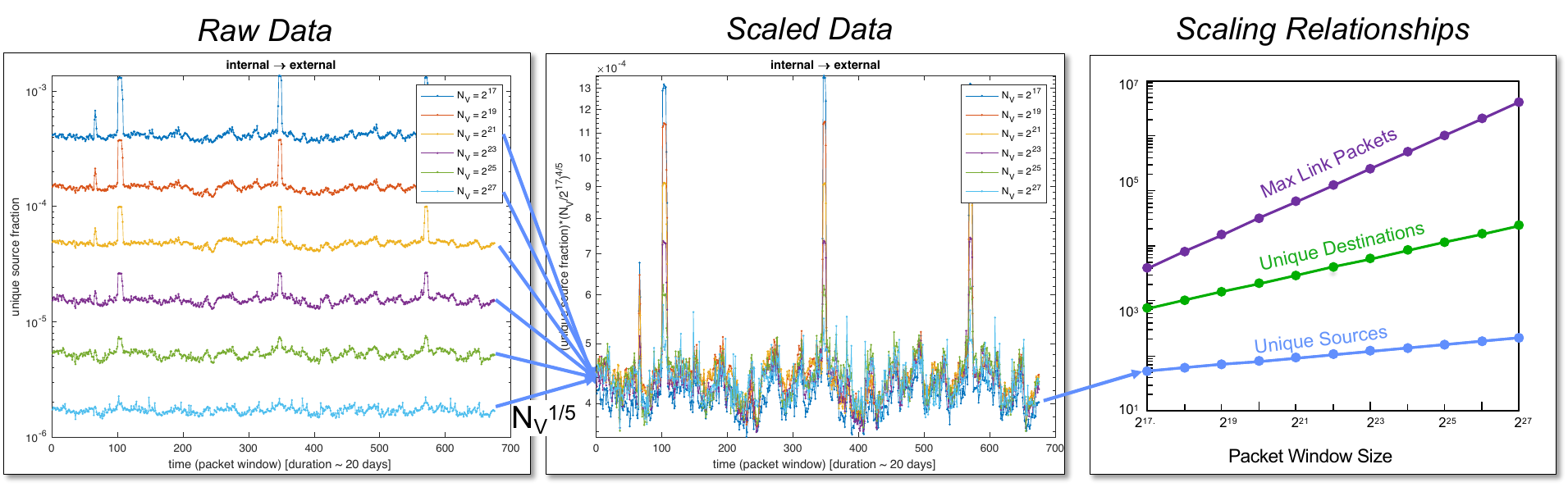}}
      	\caption{{\bf Scaling with Packet Window Size.} Network quantities vary with packet window size $N_V$. This example is derived from 100 billion packets collected at a large enterprise gateway.  ({\bf left}) Unique external sources seen over time as a fraction of total packets for different window sizes illustrating the decreasing uniqueness as window size increases from $N_V = 2^{17}$ to $2^{27}$. ({\bf middle}) Data on the left divided by $N_V^{4/5}$ indicates that the number of unique sources is proportional to $N_V/ N_V^{4/5} = N_V^{1/5}$.  ({\bf right}) Scaling of other network quantities from the same data set (see Table II in \cite{kepner2020multi}): unique sources $\approx 5{\times}N_V^{1/5}$, unique destinations $\approx 2{\times}N_V^{1/2}$, and max link packets $\approx 0.03{\times}N_V^1$.  [Note: while these scaling relationships are broadly observed the specific parameters are often site specific but stable over time \cite{kepner2020multi, kepner2021spatial}.]}
      	\label{fig:NvScaling}
\end{figure*}

The network quantities depicted in Figure~\ref{fig:NetworkDistribution} are computable from anonymized origin-destination traffic  matrices.  It is common to filter network packets down to a valid subset of packets for any particular analysis.   Such filters may limit particular sources, destinations, protocols, and time windows.   At a given time $t$, $N_V$ consecutive valid packets are aggregated from the traffic into a hypersparse matrix ${\bf A}_t$, where ${\bf A}_t(i,j)$ is the number of valid packets between the source $i$ and destination $j$. The sum of all the entries in ${\bf A}_t$ is equal to $N_V$
$$
    \sum_{i,j} {\bf A}_t(i,j) = N_V
$$
All the network quantities depicted in Figure~\ref{fig:NetworkDistribution} can be readily computed from ${\bf A}_t$ using the formulas listed in Table~\ref{tab:Aggregates}.  Because matrix operations are generally invariant to permutation (reordering of the rows and columns), these quantities can readily be computed from anonymized data.  Furthermore, the anonymized data can be analyzed by source and destination subranges (subsets when anonymized)  using simple matrix multiplication.  For a given subrange represented by an anonymized hypersparse diagonal matrix ${\bf A}_r$, where ${\bf A}_r(i,i) = 1$ implies  source/destination $i$ is in the range, the traffic within the subrange can be computed via: ${\bf A}_r {\bf A}_t  {\bf A}_r$. Likewise, for additional privacy guarantees that can be implemented at the  edge, the same method can be used to exclude a range of data from the traffic matrix
$$
     {\bf A}_t - {\bf A}_r {\bf A}_t  {\bf A}_r
$$ 

One of the important capabilities of the award-winning SuiteSparse GraphBLAS\cite{davis2019algorithm} library is direct support of hypersparse matrices where the number of nonzero entries is significantly less than either dimensions of the matrix \cite{bulucc2009parallel}.
If the packet source and destination identifiers are drawn from a large numeric range, such as those used in the Internet protocol, then a hypersparse representation of ${\bf A}_t$ eliminates the need to keep track of additional indices and can significantly accelerate the computations \cite{kepner2021vertical}.

\section{Internet Statistical Properties}

When considering what is normal in a particular domain from an observational science perspective a core question to address is 
\begin{itemize}
\item[Q~~] Given two observers at different locations and/or times what can they both expect to see?
\end{itemize}
Answering this question sets the foundation for observational reproducibility that is essential for scientific understanding.  To simplify the investigation in the specific context of Internet traffic the above question can be decomposed into several narrower questions that can be explored individually
\begin{itemize}
\item[Q1] Given a sample of $N_V$ Internet packets what are the expected values of various network quantities?
\item[Q2] What is the probability of seeing a specific value of a network quantity?
\item[Q3] Having seen a source of Internet packets what is the probability of seeing that source at a later time?
\item[Q4] What is the probability that two observers will see the same source at a given time?
\end{itemize}
Q1 deals with number of packets or the size of a packet window in a given sample of network data.  Q2 focuses on the probability distributions obtained from the histograms of the network quantities.  Q3 deals with the temporal self-correlations within a specific Internet traffic sensor while Q4 deals with the temporal cross-correlations of separate Internet traffic sensors.

Exploring these questions requires big data.  The subsequent analysis draws from the following Internet traffic data sets which are among the largest available for scientific research
\begin{itemize}
\item CAIDA Telescope: over 40 trillion mostly malicious packets collected on an Internet darkspace  over several years \cite{kepner2021spatial}
\item MAWI: several billion mostly benign packets collected at multiple sites as part the day-in-the-life of the Internet project \cite{kepner19hypersparse, kepner2022new}
\item GreyNoise: hundreds of millions of mostly malicious web interactions collected over several years from thousands of honeypot systems spread across the Internet \cite{kepner2022temporal, jananthan2023mapping}
\item Enterprise gateway: over 100 billion mostly benign packets collected at a large organization \cite{kepner2020multi}
\end{itemize}

\begin{figure*}
\center{\includegraphics[width=1.8\columnwidth]{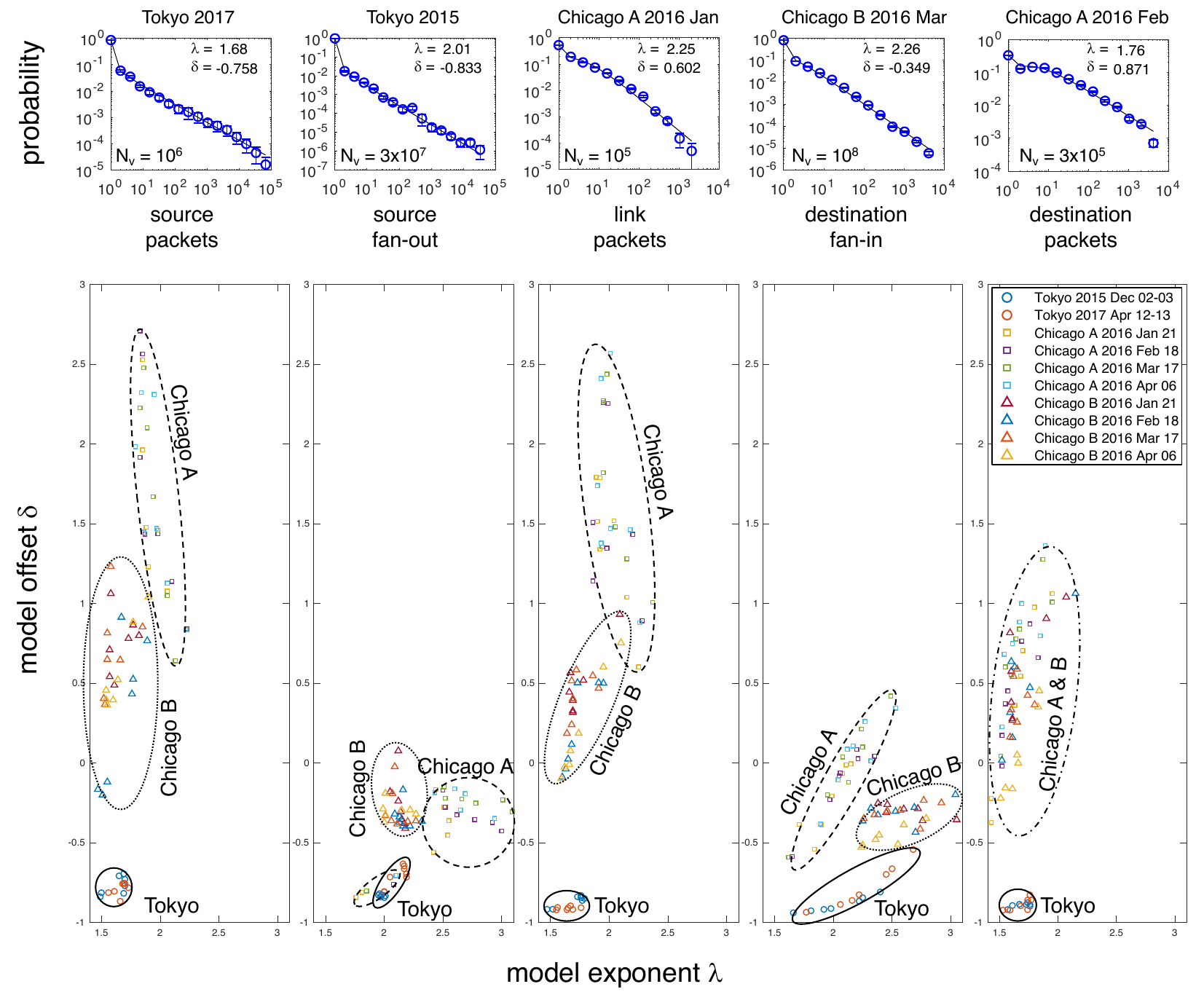}}
      	\caption{{\bf Power Law Distribution of Network Quantity Probabilities.} ({\bf top}) Probability distributions of 5 representative measured network quantities (source packets, source fan-out, link packets, destination fan-in, and destination packets) spanning different locations, dates, and packet windows from the multi-billion packet MAWI data set.  Blue circles are measured data with $\pm$1-$\sigma$ error bars.  Black lines are the best-fit modified Zipf--Mandelbrot models with parameters $\delta$ and $\lambda$.  ({\bf bottom}) Model fit parameters of the same 5 network quantities for 350 measured probability distributions for all locations, times, and sample windows sizes in the MAWI data sample; illustrating the relatively stability over time of model parameters at a given site. Figure adapted from \cite{kepner19hypersparse, kepner2022new}.}
      	\label{fig:MAWI-PowerLaw}
\end{figure*}

\subsection{Sample Window Size}

One of the first questions encountered when analyzing Internet traffic is how many samples (packets) to collect and at what level of granularity.  It is common to filter network packets down to a valid subset of packets for any particular analysis so that at a given time $t$, $N_V$ consecutive valid packet have been collected for analysis in a traffic matrix.  Statistical fluctuations between samples are significantly reduced if $N_V$ is held fixed and the sample time window is allowed to vary.

As $N_V$ increases, the network quantities in Figure~\ref{fig:NetworkDistribution} and Table~\ref{tab:Aggregates} will all increase.  How will the network quantities increase as a function of $N_V$?  For small values of  $N_V$ starting at 1 the network quantities may increase linearly.  For sufficiently large values of $N_V$ the packets may fill the entire allowed range of sources and/or destinations of the network sensor and the network quantities may level off.  Exploring this question with the various large data sets indicates that for intermediate values of $N_V$ the network quantities are often proportional to
$$
   N_V^\gamma
$$
where $0 \le \gamma \le 1$.  Figure~\ref{fig:NvScaling} illustrates a specific example derived from 100 billion packets collected at a large enterprise gateway \cite{kepner2020multi}.  These scaling relationships are broadly observed with the specific values of the parameters being site specific but stable over time \cite{kepner2020multi, kepner2021spatial}.

\begin{figure*}
\center{\includegraphics[width=1.8\columnwidth]{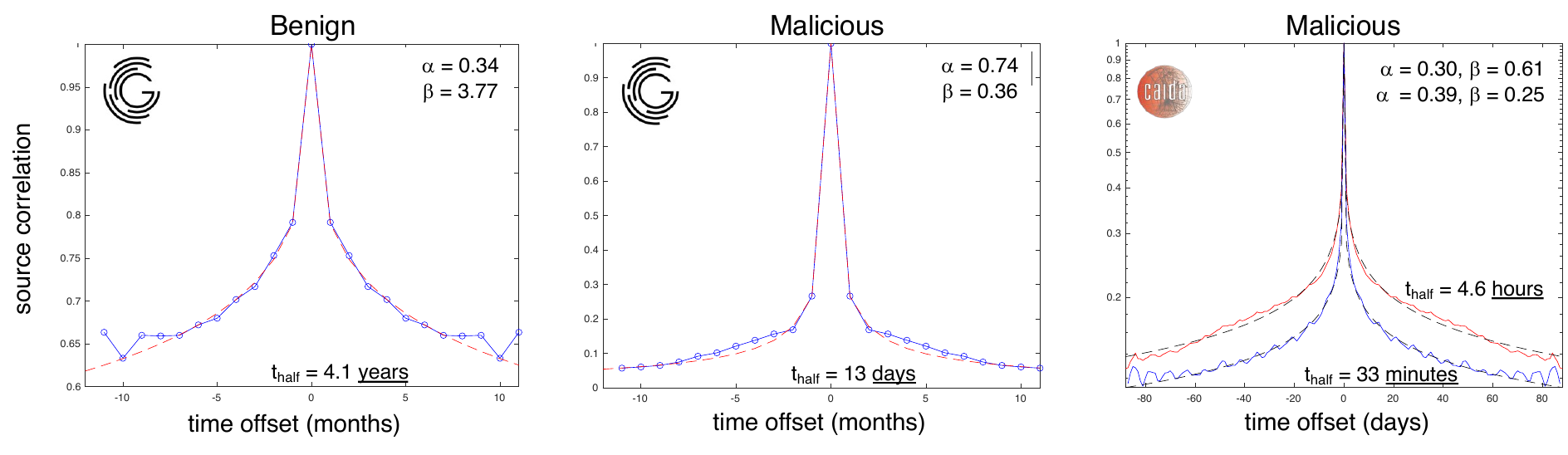}}
      	\caption{{\bf Internet Source Temporal Self-Correlations.}  ({\bf left \& middle}) Self-correlations among different categories of sources (benign and malicious ) in the GreyNoise honeyfarm from 2021Q2 thru 2022Q1. ({\bf left}) Source self-correlations among sources observed by the CAIDA darknet telescope during 2022Q1 at noon (upper curve) and midnight (lower currve).  Each point represents the sources drawn from a packet window with $N_V = 2^{30}$ valid packets.  Solid lines denote measured data.  Dashed lines correspond to the best fit modified Cauchy distribution. Corresponding modified Cauchy parameters and full-width-half-maximum time $t_{\rm half} = \beta^{1/\alpha}$ are shown illustrating the significant difference between benign and malicious traffic. Figure adapted from \cite{jananthan2023mapping}.}
      	\label{fig:SelfCorrelation}
\end{figure*}

\subsection{Probability Distributions}

    Perhaps one of the most significant early results from analyzing the Internet, which helped establish the emerging field of Network Science, was the observation that many network quantities follow a power-law or heavy-tail distribution  \cite{barabasi2016network, nair2020fundamentals}.  In terms of Internet traffic, an example would be that a few destinations on the Internet receive packets from many sources while most destinations receive packets from a few sources.  This question is readily explored by the looking at the histograms or probability distributions of network quantities computed from anonymized traffic matrices.    The availability of larger data sets have allowed the observations of these probability distributions to become more precise \cite{kepner19hypersparse, kepner2022new}.  Specifically, the probability of a particular network quantity having a value or degree $d$ is often well-described by the Zipf-Mandelbrot distribution
$$
\frac{1}{(d + \delta)^{\lambda}}
$$
where typically $-1 \lesssim \delta \lesssim 3$ and $1 \lesssim \lambda \lesssim 3$.  Given sufficient observations,  $\delta$ and $\lambda$ can be determined with high-precision.  Figure~\ref{fig:MAWI-PowerLaw} illustrates the Zipf-Mandelbrot behavior observed from billions of packets from the MAWI data set \cite{kepner19hypersparse, kepner2022new}.  Similar to the window size scaling relationships,  the Zipf-Mandelbrot distribution is broadly observed with the specific values of the parameters being site specific but stable over time.

Historically, it is worth noting that initial interest in these distributions focused on the power-law parameter $\lambda$, as this parameter described the behavior of the largest and most popular sources on the Internet \cite{barabasi2016network}.  More recently $\delta$ has emerged as a way of describing large numbers of less popular sources that may be collectively involved in adversarial network activity.

\subsection{Temporal Self-Correlations}

If an observer sees a source on the Internet what is the probability that the source will be seen again at a later time?  This is the essential question that self-correlations seek to answer.  The network traffic data sets can be used to address this questions by measuring the probability of seeing a source again at time $t$. Figure~\ref{fig:SelfCorrelation} illustrates these probabilities for the CAIDA and GreyNoise data sets over months and years \cite{jananthan2023mapping}.  Intriguingly the source self-correlations are well approximated by the modified Cauchy distribution
$$
\frac{\beta}{\beta + t^\alpha}
$$
where typically $0 < \alpha \lesssim 1$ and $\beta > 0$.  These parameters are site specific and differ significantly between benign and malicious data.  The modified Cauchy distribution can be characterized by the time it takes for the probability to drop to one half
$$
  t_{\rm half} = \beta^{1/\alpha}
$$
In the case of the GreyNoise benign data this timescale is years while the malicious data has much shorter times scales of days, hours, and minutes.

\subsection{Temporal Cross-Correlations}

\begin{figure}
\center{\includegraphics[width=1.0\columnwidth]{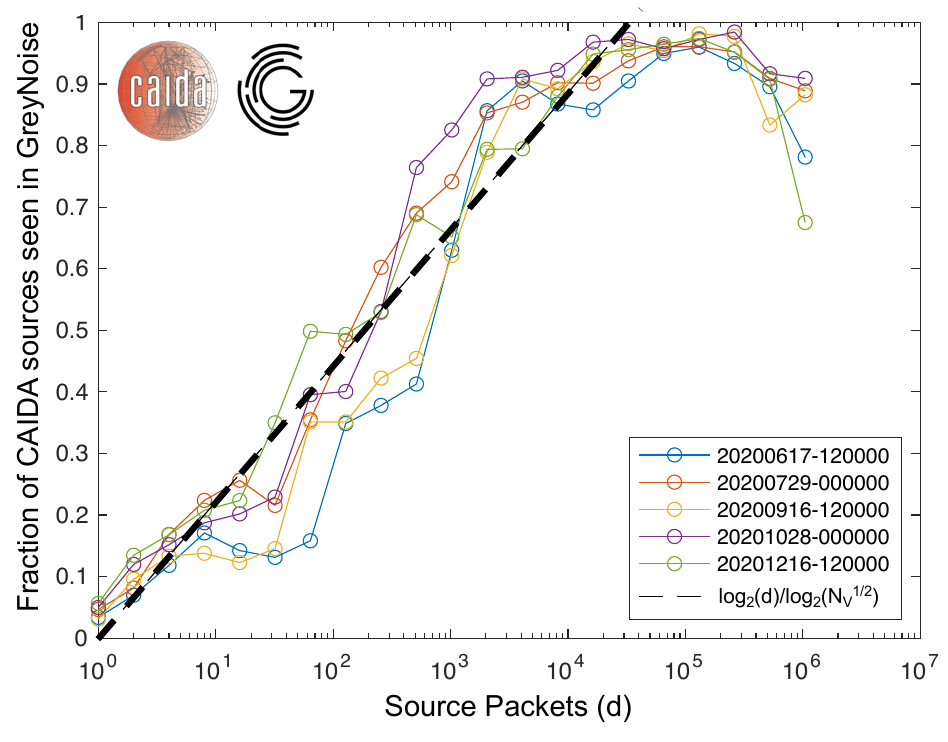}}
      	\caption{{\bf Internet Source Cross-Correlations.}  Correlation of CAIDA sources with GreyNoise sources during the same month as a function of CAIDA source packets $d$.  CAIDA sources with $d > N_V^{1/2}$ packets in the packet window ($N_V = 2^{30}$) are very likely to appear in the GreyNoise data of the same month. CAIDA sources with $d < N_V^{1/2}$ appear with a probability  $\log_2(d)/\log_2(N_V^{1/2})$. Figure adapted from \cite{kepner2022temporal}.}
      	\label{fig:CrossCorrelation}
\end{figure}

Similar to self-correlations, it is likewise possible to explore the probability that a second observer will see a source at the same or a different time.  If the self-correlations of two network sensors are observed to follow a modified Cauchy distribution it is not surprising that their cross-correlations in time are also observed to follow a modified Cauchy distribution \cite{kepner2022temporal}.   Perhaps more fundamental is the probability that a source seen by one observer will even be seen by another observer.  Figure~\ref{fig:CrossCorrelation} plots the probability of a source seen by the CAIDA telescope also being seen in the same month by the GreyNoise honeyfarm.  This probability is strongly dependent upon the number of packets the source has sent to the CAIDA telescope and is well approximated by the formula
$$
  \frac{\log_2(d)}{\log_2(N_V^{1/2})}
$$
for $d < N_V^{1/2}$.  Simply put, if a source emits a lot of packets it is more likely to be seen.

\section{Model Synthesis}

The empirically motivated models from the previous section allow the core question to be refined around the variables that directly impact the observability of network traffic.  These variables include the window size $N_V$, the number of packets $d$ observed from a source, and the time $t$ between observations. The more pricess question then becomes  
\begin{itemize}
\item[\underline{Q}] Given a window with $N_V$ incoming packets, what is the probability of a source sending $d$ packets being observed by a second observer at time $t$
\end{itemize}
Based on prior observations, the empirical formula for this probability can be hypothesized to be proportional to
$$
N_V^\gamma ~ 
\frac{1}{(d + \delta)^{\lambda}} ~
\frac{\beta}{\beta + t^{\alpha}} ~
\frac{\log_2(d)}{\log_2(N_V^{1/2})}
$$
where $\gamma$, $\delta$, $\lambda$, $\beta$, and $\alpha$ are site specific parameters that tend to be stable over time.

The above expression is an empirically motivated formula.  Ideally, theoretical models derived from underlying first-principles will be found which are then approximated by the above formula under appropriate conditions.  While such a theoretical model does not yet exist, certain logical deductions can be made about the reasonableness of the terms in the above formula.

By definition network quantities grow monotonically with window size and the term $N_V^\gamma$ is one the simplest formulas satisfying this condition.  The power-law dependence of network quantities on their observed value $d$ has been well-observed and the successful preferential attachment model remains a reasonable underlying theoretical framework for these observations \cite{barabasi2016network}.  Temporal self-correlation and cross-correlation measurements require continuous long-duration coeval observations from multiple observers and are subsequently rarer.  The correlation function is defined to have a peak value of 1 at $t=0$ and it seems intuitive that the probability of seeing a source again would slowly drop-off over time.  The modified Cauchy distribution satisfies these conditions. Finally, it is also intuitive that the probability of an observer seeing a source is related to the ratio of the number of packets from the source and the size of the observation window.

\section{Conclusions and Future Work}

Modern Internet telescopes and high-performance sensors are capable of collecting trillions of observations.  Supercomputers and high performance graph analysis libraries, such as the GraphBLAS, allow these big data observations to be analyzed to develop high-precision, low-parameter observational models of Internet traffic.  These models provide detailed predictions on the visibility of Internet sources of a given intensity over time and the likelihood such sources will be seen by an observer at a different location.  For a given location the parameters of the model tend to be stable over time.  Using these models, it is possible to predict in detail many statistical properties of Internet traffic seen at a given location and time.  These predictions can assist in correctly placing network sensors by comparing what is expected with what is observed, ensuring zero trust configurations are maintained by revealing when networks have changed, and detecting anomalies due to malicious activity.

Going forward there should be an expansion of Internet observatories like CAIDA and MAWI.  The globe currently depends upon a small dedicated community to operate and maintain current network observatories. These lookouts are our only means for obtaining consensus empirical answers to critical questions. These capabilities should be significantly expanded.  Furthermore, the underlying network science at scale needs enhancement.  Understanding of the underlying processes in any field is discovered by painstaking science. Early efforts on small data sets revealed significant new discoveries and established the field of Network Science \cite{barabasi2016network}. Current observations are a much larger and are calling out for scientific exploration.

\section*{Acknowledgments}

The authors wish to acknowledge the following individuals for their contributions and support: 
Daniel Andersen, LaToya Anderson, Sean Atkins, Chris Birardi, Bob Bond, Alex Bonn, Koley Borchard, Stephen Buckley, Aydin Buluc, K Claffy, Cary Conrad, Chris Demchak, Phil Dykstra, Alan Edelman, Garry Floyd, Jeff Gottschalk, Dhruv Gupta, Thomas Hardjono, Chris Hill, Charles Leiserson, Kirsten Malvey, Chad Meiners, Adam Michaleas, Sanjeev Mohindra, Heidi Perry, Christian Prothmann, Steve Rejto, Josh Rountree, Daniela Rus, Mark Sherman, Scott Weed, Adam Wierman, Marc Zissman.

\bibliographystyle{ieeetr}
\bibliography{TrafficModelling}

\begin{thebibliography}{10}

\bibitem{ahmed2016survey}
M.~Ahmed, A.~N. Mahmood, and J.~Hu, ``A survey of network anomaly detection
  techniques,'' {\em Journal of Network and Computer Applications}, vol.~60,
  pp.~19--31, 2016.

\bibitem{fernandes2019comprehensive}
G.~Fernandes, J.~J. Rodrigues, L.~F. Carvalho, J.~F. Al-Muhtadi, and M.~L.
  Proen{\c{c}}a, ``A comprehensive survey on network anomaly detection,'' {\em
  Telecommunication Systems}, vol.~70, pp.~447--489, 2019.

\bibitem{kwon2019survey}
D.~Kwon, H.~Kim, J.~Kim, S.~C. Suh, I.~Kim, and K.~J. Kim, ``A survey of deep
  learning-based network anomaly detection,'' {\em Cluster Computing}, vol.~22,
  pp.~949--961, 2019.

\bibitem{singla2019deep}
A.~Singla and E.~Bertino, ``How deep learning is making information security
  more intelligent,'' {\em IEEE Security \& Privacy}, vol.~17, no.~3,
  pp.~56--65, 2019.

\bibitem{kaur2020hybrid}
S.~Kaur and M.~Singh, ``Hybrid intrusion detection and signature generation
  using deep recurrent neural networks,'' {\em Neural Computing and
  Applications}, vol.~32, no.~12, pp.~7859--7877, 2020.

\bibitem{wang2021machine}
S.~Wang, J.~F. Balarezo, S.~Kandeepan, A.~Al-Hourani, K.~G. Chavez, and
  B.~Rubinstein, ``Machine learning in network anomaly detection: A survey,''
  {\em IEEE Access}, vol.~9, pp.~152379--152396, 2021.

\bibitem{chandola2009anomaly}
V.~Chandola, A.~Banerjee, and V.~Kumar, ``Anomaly detection: A survey,'' {\em
  ACM computing surveys (CSUR)}, vol.~41, no.~3, pp.~1--58, 2009.

\bibitem{watson1957three}
R.~A. Watson-Watt, {\em Three Steps to Victory: A Personal Account by Radar's
  Greatest Pioneer}.
\newblock London: Odhams Press, 1957.

\bibitem{delaney1990air}
W.~P. Delaney, ``Air defense of the united states: Strategic missions and
  modern technology,'' {\em International Security}, vol.~15, no.~1,
  pp.~181--211, 1990.

\bibitem{delaney2015perspectives}
W.~P. Delaney, {\em Perspectives on Defense Systems Analysis}.
\newblock MIT Press, 2015.

\bibitem{geul2017modelling}
J.~Geul, E.~Mooij, and R.~Noomen, ``Modelling and assessment of the current and
  future space surveillance network,'' {\em 7th ECSD}, 2017.

\bibitem{o2018radar}
K.~W. O’Haver, C.~K. Barker, G.~D. Dockery, and J.~D. Huffaker, ``Radar
  development for air and missile defense,'' {\em Johns Hopkins APL Tech.
  Digest}, vol.~34, no.~2, pp.~140--153, 2018.

\bibitem{pisharody2021realizing}
S.~Pisharody, J.~Bernays, V.~Gadepally, M.~Jones, J.~Kepner, C.~Meiners,
  P.~Michaleas, A.~Tse, and D.~Stetson, ``Realizing forward defense in the
  cyber domain,'' in {\em 2021 IEEE High Performance Extreme Computing
  Conference (HPEC)}, pp.~1--7, IEEE, 2021.

\bibitem{jason2010science}
JASON, ``{Science of Cyber-Security},'' Tech. Rep. JSR-10-102, The MITRE
  Corporation, McLean, VA, Nov. 2010.

\bibitem{carroll2012realizing}
T.~E. Carroll, D.~Manz, T.~Edgar, and F.~L. Greitzer, ``Realizing scientific
  methods for cyber security,'' in {\em Proceedings of the 2012 Workshop on
  Learning from Authoritative Security Experiment Results}, pp.~19--24, 2012.

\bibitem{thuraisingham2016data}
B.~Thuraisingham, M.~Kantarcioglu, K.~Hamlen, L.~Khan, T.~Finin, A.~Joshi,
  T.~Oates, and E.~Bertino, ``A data driven approach for the science of cyber
  security: Challenges and directions,'' in {\em 2016 IEEE 17th International
  Conference on Information Reuse and Integration (IRI)}, pp.~1--10, IEEE,
  2016.

\bibitem{spring2017practicing}
J.~M. Spring, T.~Moore, and D.~Pym, ``Practicing a science of security: A
  philosophy of science perspective,'' in {\em Proceedings of the 2017 New
  Security Paradigms Workshop}, pp.~1--18, 2017.

\bibitem{leland1994self}
W.~E. Leland, M.~S. Taqqu, W.~Willinger, and D.~V. Wilson, ``On the
  self-similar nature of ethernet traffic (extended version),'' {\em IEEE/ACM
  Transactions on Networking (ToN)}, vol.~2, no.~1, pp.~1--15, 1994.

\bibitem{faloutsos1999power}
M.~Faloutsos, P.~Faloutsos, and C.~Faloutsos, ``On power-law relationships of
  the internet topology,'' in {\em ACM SIGCOMM computer communication review},
  vol.~29-4, pp.~251--262, ACM, 1999.

\bibitem{barabasi1999emergence}
A.-L. Barab{\'a}si and R.~Albert, ``Emergence of scaling in random networks,''
  {\em Science}, vol.~286, no.~5439, pp.~509--512, 1999.

\bibitem{albert1999internet}
R.~Albert, H.~Jeong, and A.-L. Barab{\'a}si, ``Internet: Diameter of the
  world-wide web,'' {\em Nature}, vol.~401, no.~6749, p.~130, 1999.

\bibitem{clauset2009power}
A.~Clauset, C.~R. Shalizi, and M.~E. Newman, ``Power-law distributions in
  empirical data,'' {\em SIAM review}, vol.~51, no.~4, pp.~661--703, 2009.

\bibitem{mahanti2013tale}
A.~Mahanti, N.~Carlsson, A.~Mahanti, M.~Arlitt, and C.~Williamson, ``A tale of
  the tails: Power-laws in internet measurements,'' {\em IEEE Network},
  vol.~27, no.~1, pp.~59--64, 2013.

\bibitem{barabasi2016network}
A.-L. Barab{\'a}si {\em et~al.}, {\em Network science}.
\newblock Cambridge university press, 2016.

\bibitem{cao18impact}
Z.~Cao, Z.~He, and N.~F. Johnson, ``Impact on the topology of power-law
  networks from anisotropic and localized access to information,'' {\em Phys.
  Rev. E}, vol.~98, p.~042307, Oct 2018.

\bibitem{fan2004prefix}
J.~Fan, J.~Xu, M.~H. Ammar, and S.~B. Moon, ``Prefix-preserving ip address
  anonymization: measurement-based security evaluation and a new
  cryptography-based scheme,'' {\em Computer Networks}, vol.~46, no.~2,
  pp.~253--272, 2004.

\bibitem{kepner2021zero}
J.~Kepner, J.~Bernays, S.~Buckley, K.~Cho, C.~Conrad, L.~Daigle, K.~Erhardt,
  V.~Gadepally, B.~Greene, M.~Jones, R.~Knake, B.~Maggs, P.~Michaleas,
  C.~Meiners, A.~Morris, A.~Pentland, S.~Pisharody, S.~Powazek, A.~Prout,
  P.~Reiner, K.~Suzuki, K.~Takhashi, T.~Tauber, L.~Walker, and D.~Stetson,
  ``Zero botnets: An observe-pursue-counter approach.'' Belfer Center Reports,
  6 2021.

\bibitem{uhlig2006providing}
S.~Uhlig, B.~Quoitin, J.~Lepropre, and S.~Balon, ``Providing public intradomain
  traffic matrices to the research community,'' {\em ACM SIGCOMM Computer
  Communication Review}, vol.~36, no.~1, pp.~83--86, 2006.

\bibitem{tune2013internet}
P.~Tune, M.~Roughan, H.~Haddadi, and O.~Bonaventure, ``Internet traffic
  matrices: A primer,'' {\em Recent Advances in Networking}, vol.~1, pp.~1--56,
  2013.

\bibitem{vinayakumar2017applying}
R.~Vinayakumar, K.~Soman, and P.~Poornachandran, ``Applying deep learning
  approaches for network traffic prediction,'' in {\em 2017 International
  Conference on Advances in Computing, Communications and Informatics
  (ICACCI)}, pp.~2353--2358, IEEE, 2017.

\bibitem{davis2019algorithm}
T.~A. Davis, ``Algorithm 1000: Suitesparse: Graphblas: Graph algorithms in the
  language of sparse linear algebra,'' {\em ACM Transactions on Mathematical
  Software (TOMS)}, vol.~45, no.~4, pp.~1--25, 2019.

\bibitem{gadepally2018hyperscaling}
V.~{Gadepally}, J.~{Kepner}, L.~{Milechin}, W.~{Arcand}, D.~{Bestor},
  B.~{Bergeron}, C.~{Byun}, M.~{Hubbell}, M.~{Houle}, M.~{Jones},
  P.~{Michaleas}, J.~{Mullen}, A.~{Prout}, A.~{Rosa}, C.~{Yee}, S.~{Samsi}, and
  A.~{Reuther}, ``Hyperscaling internet graph analysis with d4m on the mit
  supercloud,'' in {\em 2018 IEEE High Performance extreme Computing Conference
  (HPEC)}, pp.~1--6, Sep. 2018.

\bibitem{kepner19hypersparse}
J.~{Kepner}, K.~{Cho}, K.~{Claffy}, V.~{Gadepally}, P.~{Michaleas}, and
  L.~{Milechin}, ``Hypersparse neural network analysis of large-scale internet
  traffic,'' in {\em 2019 IEEE High Performance Extreme Computing Conference
  (HPEC)}, pp.~1--11, 2019.

\bibitem{kepner2020multi}
J.~Kepner, C.~Meiners, C.~Byun, S.~McGuire, T.~Davis, W.~Arcand, J.~Bernays,
  D.~Bestor, W.~Bergeron, V.~Gadepally, R.~Harnasch, M.~Hubbell, M.~Houle,
  M.~Jones, A.~Kirby, A.~Klein, L.~Milechin, J.~Mullen, A.~Prout, A.~Reuther,
  A.~Rosa, S.~Samsi, D.~Stetson, A.~Tse, C.~Yee, and P.~Michaleas,
  ``Multi-temporal analysis and scaling relations of 100,000,000,000 network
  packets,'' in {\em 2020 IEEE High Performance Extreme Computing Conference
  (HPEC)}, pp.~1--6, 2020.

\bibitem{kepner2021spatial}
J.~Kepner, M.~Jones, D.~Andersen, A.~Buluç, C.~Byun, K.~Claffy, T.~Davis,
  W.~Arcand, J.~Bernays, D.~Bestor, W.~Bergeron, V.~Gadepally, M.~Houle,
  M.~Hubbell, A.~Klein, C.~Meiners, L.~Milechin, J.~Mullen, S.~Pisharody,
  A.~Prout, A.~Reuther, A.~Rosa, S.~Samsi, D.~Stetson, A.~Tse, C.~Yee, and
  P.~Michaleas, ``Spatial temporal analysis of 40,000,000,000,000 internet
  darkspace packets,'' in {\em 2021 IEEE High Performance Extreme Computing
  Conference (HPEC)}, pp.~1--8, 2021.

\bibitem{bulucc2009parallel}
A.~Bulu{\c{c}}, J.~T. Fineman, M.~Frigo, J.~R. Gilbert, and C.~E. Leiserson,
  ``Parallel sparse matrix-vector and matrix-transpose-vector multiplication
  using compressed sparse blocks,'' in {\em Proceedings of the twenty-first
  annual symposium on Parallelism in algorithms and architectures},
  pp.~233--244, 2009.

\bibitem{kepner2021vertical}
J.~Kepner, T.~Davis, C.~Byun, W.~Arcand, D.~Bestor, W.~Bergeron, V.~Gadepally,
  M.~Houle, M.~Hubbell, M.~Jones, A.~Klein, L.~Milechin, J.~Mullen, A.~Prout,
  A.~Reuther, A.~Rosa, S.~Samsi, C.~Yee, and P.~Michaleas, ``Vertical,
  temporal, and horizontal scaling of hierarchical hypersparse graphblas
  matrices,'' in {\em 2021 IEEE High Performance Extreme Computing Conference
  (HPEC)}, pp.~1--6, 2021.

\bibitem{kepner2022new}
J.~Kepner, K.~Cho, K.~Claffy, V.~Gadepally, S.~McGuire, L.~Milechin, W.~Arcand,
  D.~Bestor, W.~Bergeron, C.~Byun, M.~Hubbell, M.~Houle, M.~Jones, A.~Prout,
  A.~Reuther, A.~Rosa, S.~Samsi, C.~Yee, and P.~Michaleas, ``New phenomena in
  large-scale internet traffic,'' in {\em Massive Graph Analytics} (D.~Bader,
  ed.), pp.~1--53, Chapman and Hall/CRC, 2022.

\bibitem{kepner2022temporal}
J.~Kepner, M.~Jones, D.~Andersen, A.~Buluc, C.~Byun, K.~Claffy, T.~Davis,
  W.~Arcand, J.~Bernays, D.~Bestor, W.~Bergeron, V.~Gadepally, D.~Grant,
  M.~Houle, M.~Hubbell, H.~Jananthan, A.~Klein, C.~Meiners, L.~Milechin,
  A.~Morris, J.~Mullen, S.~Pisharody, A.~Prout, A.~Reuther, A.~Rosa, S.~Samsi,
  D.~Stetson, C.~Yee, and P.~Michaleas, ``Temporal correlation of internet
  observatories and outposts,'' in {\em 2022 IEEE International Parallel and
  Distributed Processing Symposium Workshops (IPDPSW)}, pp.~247--254, 2022.

\bibitem{jananthan2023mapping}
H.~Jananthan, J.~Kepner, M.~Jones, W.~Arcand, D.~Bestor, W.~Bergeron, C.~Byun,
  T.~Davis, V.~Gadepally, D.~Grant, M.~Houle, M.~Hubbell, A.~Klein,
  L.~Milechin, G.~Morales, A.~Morris, J.~Mullen, R.~Patel, A.~Pentland,
  S.~Pisharody, A.~Prout, A.~Reuther, A.~Rosa, S.~Samsi, T.~Trigg, G.~Wachman,
  C.~Yee, and P.~Michaleas, ``Mapping of internet “coastlines” via large
  scale anonymized network source correlations,'' in {\em 2023 IEEE High
  Performance Extreme Computing Conference (HPEC)}, pp.~1--9, 2023.

\bibitem{nair2020fundamentals}
J.~Nair, A.~Wierman, and B.~Zwart, ``The fundamentals of heavy tails:
  Properties, emergence, and estimation,'' {\em Preprint, California Institute
  of Technology}, 2020.

\end{thebibliography}

\end{document}